\documentclass{icrc29}
\usepackage{graphicx,amssymb,amsmath,times}
\begin{document}
\title[MC Studies of the first VERITAS telescope...]{Monte Carlo Studies of the first VERITAS telescope}
\author[Gernot Maier et al.] {G.Maier$^a$ for the VERITAS Collaboration$^b$ \\
        (a) University of Leeds, LS2 9JT Leeds, United Kingdom \\
        (b) for full collaboration list see J.Holder, "Status and performance of the first VERITAS telescope", this proceedings.
        }
\presenter{Presenter: Gernot Maier (gm@ast.leeds.ac.uk), \  
uki-maier-G-abs2-og27-oral}

\maketitle

\begin{abstract}

The Very High Energy Radiation Telescope Array (VERITAS) is a system
of four imaging Cherenkov telescopes currently under construction at
Kitt Peak, Arizona, USA. The first telescope has been in operation at
the Mt. Hopkins basecamp since January 2005.
We present here detailed Monte Carlo simulations of the telescope response
to extensive air showers.
The energy threshold for this stand-alone telescope 
is calculated to be 150 GeV at trigger level, the
$\gamma$-ray trigger rate is 22 $\gamma$'s/min.
Image parameter distributions, 
and the quality of gamma-hadron discrimination are calculated
and show good agreement with distributions from
observations of background cosmic rays
and high-energy gamma-rays from the Crab Nebula and Markarian 421.
The energy spectrum of the Crab is reconstructed as
$(3.26\pm0.9)\cdot 10^{-7}\cdot E^{-(2.6\pm 0.3)}$  m$^{-2}$s$^{-1}$TeV$^{-1}$.
\end{abstract}

\vspace{-0.35cm}
\section{Introduction}

The first of the four VERITAS telescopes has been in routine operation since January 2005
at the basecamp
of the Fred Lawrence Whipple Observatory at Mt. Hopkins, Arizona.
The site is 1275 m above sea level.
Several TeV gamma ray sources have been detected \cite{JH05} and the construction of the other
telescopes is underway.
A detailed description of the first VERITAS telescope, called {\it Telescope 1} in the 
following, can be found in \cite{JH05}, 
a full description of the VERITAS project in \cite{Weekes02}.

\vspace{-0.40cm}
\section{Data}

For the following comparison of data with Monte Carlo simulations, gamma-ray candidates are
extracted from observations of the Crab Nebula (3.9 h on source) and Mrk 421 (4.2h on source)
in March and April 2005.
The data is taken in ON-OFF mode at elevations above 60$^{\circ}$ and in good weather conditions.
Images are parameterized using standard second moment analysis \cite{Hillas:1985}.
Image pixels are selected by a two level cut on the charge per pixel.
Picture and boundary thresholds are pedestal plus four respective two times
the pedestal variance.
The cuts on image parameters to select gamma-rays 
are $\alpha<10^{\circ}$, length/size $<0.002$, length $>0.14^{\circ}$, $0.067^{\circ}<$ width $<0.1^{\circ}$, and
 $0.5^{\circ}< $ distance $<1^{\circ}$.
Alpha is the orientation angle of the image ellipse relative to the direction of the
source position.
Length and width describe the shape of the image, and distance the location of the image centroid with
respect to the source position.
Only events with at least five image pixels are taken into account.
These cuts results in a combined significance of about 21.7$\sigma$ and roughly 900 gamma-ray candidates
for the observations of the Crab Nebula and Mrk 421.
The Crab data set is statistically limited due to the fact that the observation started very late
in the season.

\vspace{-0.40cm}
\section{Monte Carlo simulations}

A complete chain of Monte Carlo simulations has been developed.
It consists of air shower simulations with CORSIKA \cite{Heck} 
and a detailed simulation of the telescope response \cite{LeBohec}.
CORSIKA version 6.20 is used with the hadronic interaction models QGSJet for primary energies above 
500 GeV and FLUKA below.
Simulated hydrogen, helium, and oxygen nuclei induced air showers 
are produced in an energy range from 10 GeV (150 GeV for oxygen) to 50 TeV at different elevations.
Spectral indices are taken from fits to balloon measurements \cite{Hoerandel}.
The shower cores are distributed randomly on a circular area with a radius of 1000m around the telescope.
The isotropic distribution of the cosmic ray incident angles is simulated by randomizing the shower directions
in a cone of radius 3.5$^{\circ}$ around the pointing direction of the telescope.
Primary gamma-rays are simulated with a Crab-like spectrum in an energy range from 10 GeV to 10 TeV 
on a circular area at the ground with radius 450m.
Scatter areas are chosen according to the lateral Cherenkov light distributions of hadronic and gamma-ray showers.


Measurements of the atmospheric properties at the current site of Telescope 1 and the final one at
Kitt Peak are currently in progress.
The calculations described here use the U.S. standard atmosphere, which does not always
reflect the properties of the atmosphere in southern Arizona.
Atmospheric extinction values are estimated by MODTRAN 4 \cite{Modtran} assuming 50 km visibility at 
550 nm at ground level.
The photo-electron rate per PMT is measured to be 100-200 MHz.
This corresponds to a  night sky background rate of
$2.8 \times 10^{12}$ photons $\cdot$ m$^{-2}$ s$^{-1}$ sr$^{-1}$, which is used in the simulations.

The telescope simulations consist of two parts, the propagation of Cherenkov photons through
the optical system and the response of the camera and electronics.
The geometrical properties of the optical system is fully implemented in the simulation.
Misalignment of the mirrors and their surface roughness is taken into account.
The camera configuration is that of April 2005, i.e.~a 499 pixel camera with 1 1/8'' phototubes without
light cones.
The field of view of the camera is 3.5$^{\circ}$.
The response of the PMTs to single photons has been measured;
the single photo-electron pulse has a rise time of 3.3 ns and a width of 6.5 ns \cite{JH05}.
In the simulations a signal in a PMT is created by adding up single photo-electron pulses
with appropriate amplitude and time jitters applied.
Electronic noise and all efficiencies, including mirror reflectivities, geometrical, quantum,
and collection efficiencies, and losses due to signal transmission have been modeled.
The pulses are digitized into 2ns samples with a trace length of 24 samples reflecting the properties of the FADC system.
The trigger simulation utilizes a simplified model of the constant fraction discriminator, which is the
first stage of the VERITAS multi-level trigger, and a full realization of the pattern trigger,
requiring three adjacent pixels above threshold in a time window of 5 ns.
The currently used trigger threshold of 70 mV corresponds to about 6.7 photoelectrons.
The output of the telescope simulations, i.e.~FADC traces for all PMTs, are written to disk in the VERITAS 
raw data format.
The analysis steps, which include pedestal calculation, image cleaning, image parametrisation, and source reconstruction,
are exactly the same for simulated and real data.

\vspace{-0.40cm}
\section{Comparison of Data with Monte Carlo simulation}

\begin{figure}[ht]
\begin{center}
\includegraphics*[width=0.41\textwidth]{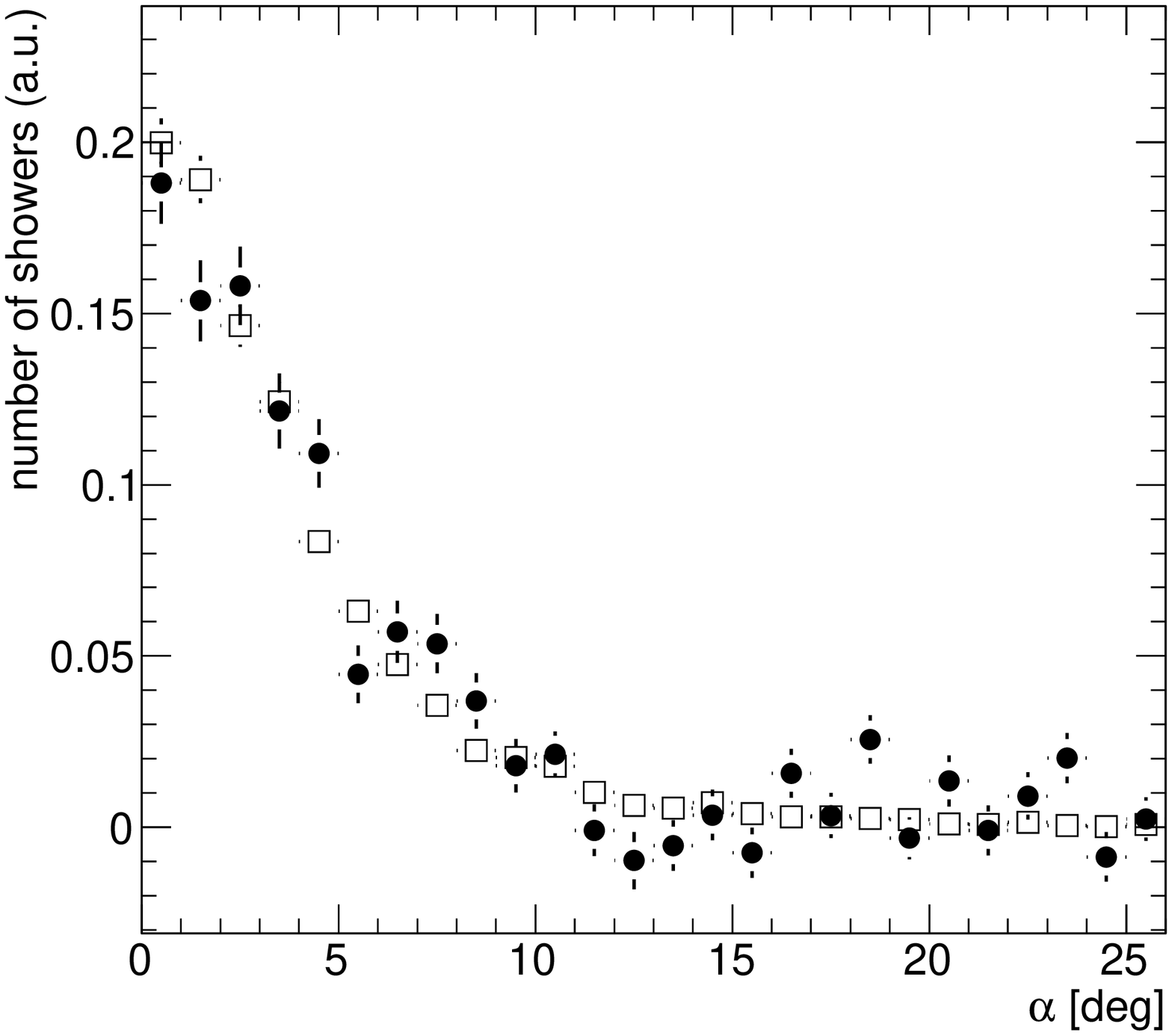}
\includegraphics*[width=0.41\textwidth]{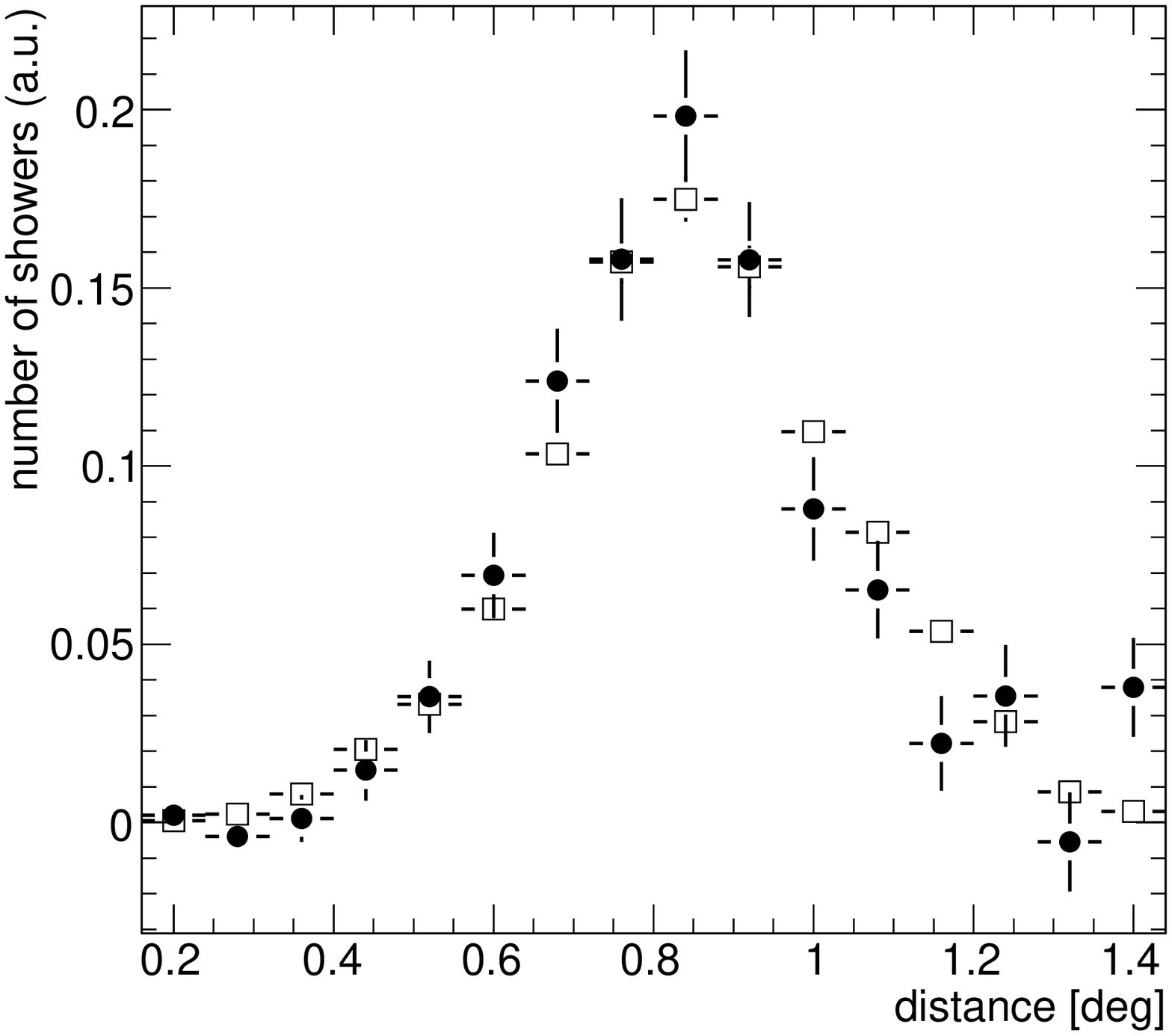}
\includegraphics*[width=0.41\textwidth]{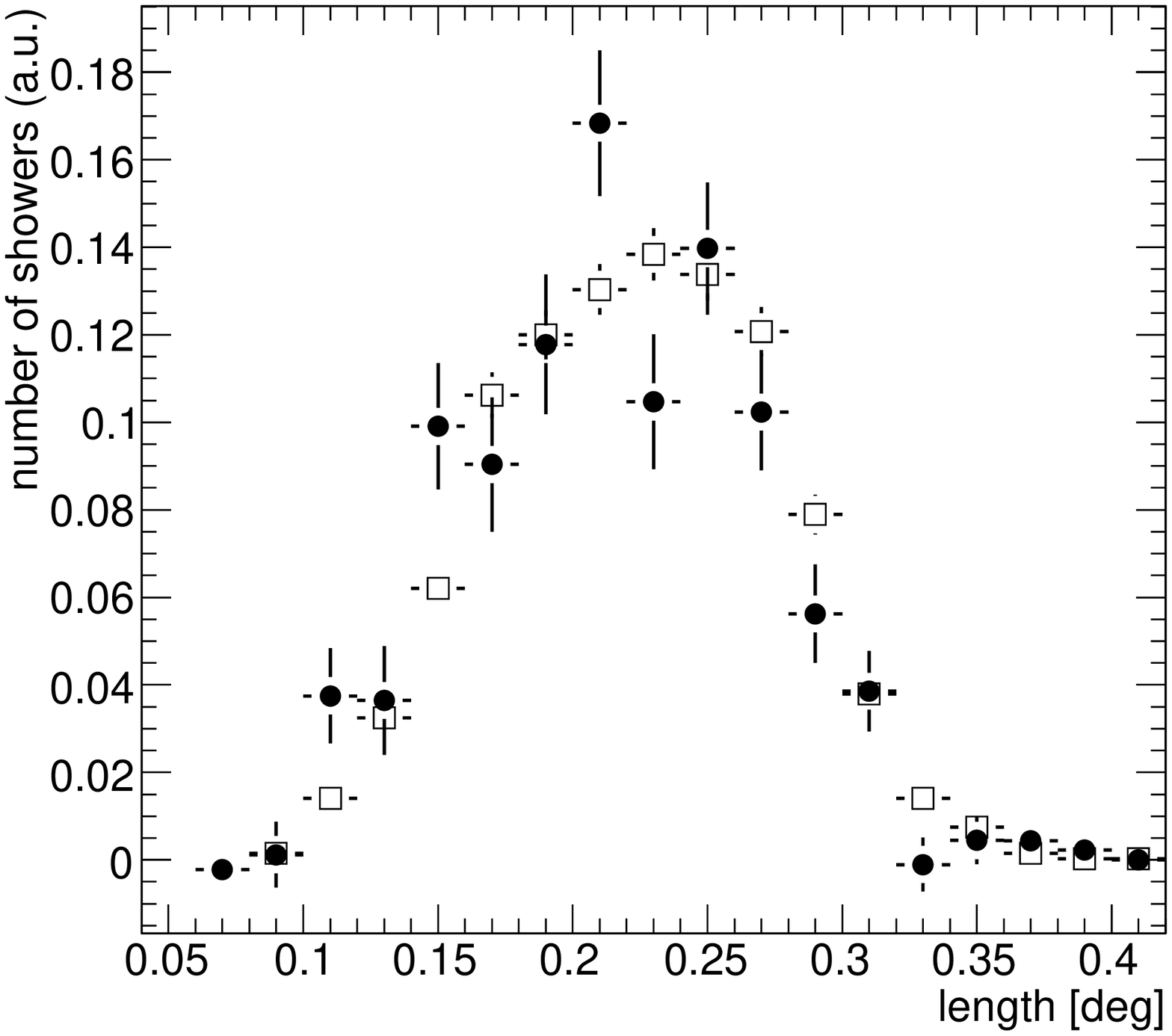}
\includegraphics*[width=0.41\textwidth]{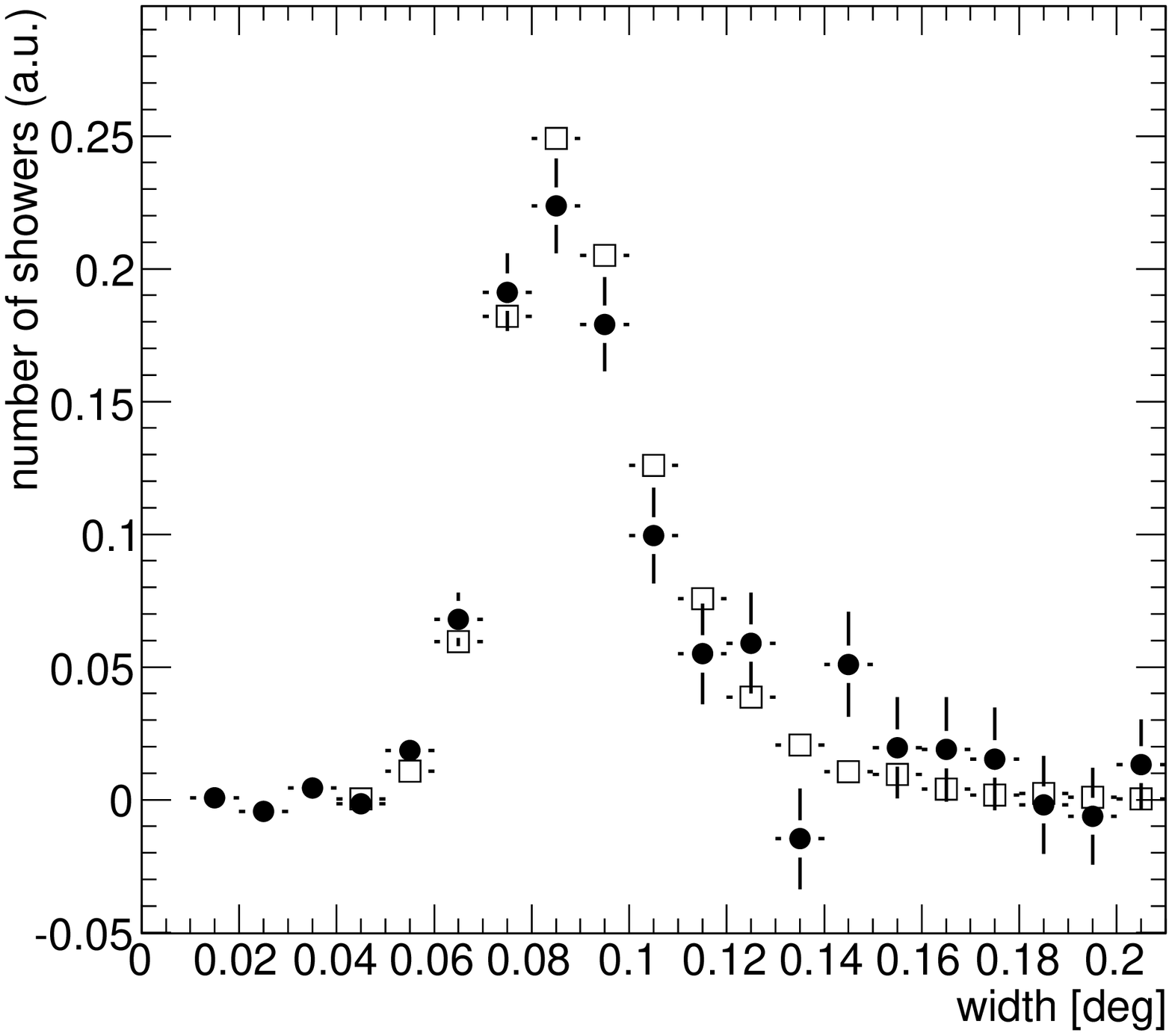}
\caption{\label {fig1} Comparison of image parameter distributions from observation data (closed symbols)
and Monte Carlo calculations (open symbols).}
\end{center}
\end{figure}

\vspace{-0.30cm}
\subsection{Trigger rate}
\vspace{-0.05cm}

The raw trigger rate of Telescope 1 with trigger conditions as described above is between 140 and 150 Hz,
mainly depending on elevation.
Dead time losses due to readout are $\sim$10\%, the corrected trigger rate is consequentially $\sim$150-160 Hz.
The simulation of the cosmic ray background results in trigger rates of 101 Hz from
air showers induced by protons, 
26 Hz from helium nuclei, and $\sim$5 Hz from nuclei of the CNO group for an elevation of 70$^{\circ}$.
The Monte Carlo calculations reproduce the observed trigger rate with an accuracy of $\sim$15\%, which is
acceptable, taking into account 
an estimated uncertainty in the cosmic ray fluxes of $\sim$25\%  and various 
systematic uncertainties in the modeling of the telescope.

\vspace{-0.40cm}
\subsection{Image parameter distribution}

Second moment image parameters are the standard tool in analyzing data from IACTs.
The suppression of the hadronic and muonic background as well as the estimation of the
primary energy depends strongly on the correct predictions of the image parameters from 
Monte Carlo calculations.
Fig.\ref{fig1} shows a comparison of image parameter distributions from real data
and simulations.
The alpha, distance, length, and width distributions show very good agreement
and proof the validity of the simulations.

The described cuts on image parameters are chosen to extract a sample of gamma ray candidates with
very little contamination from background events.
About 10\% of all gamma ray showers are accepted with these cuts, rejecting $\sim$99.9\% of all 
cosmic ray showers.
These hard cuts are necessary for data from a single telescope
to reject the large background due to local muons.
(Telescope arrays are insensitive to such single muons and can be run and analyzed with looser cuts.)

\vspace{-0.40cm}
\subsection{Performance}

\begin{figure}[h]
\begin{center}
\includegraphics*[width=0.41\textwidth]{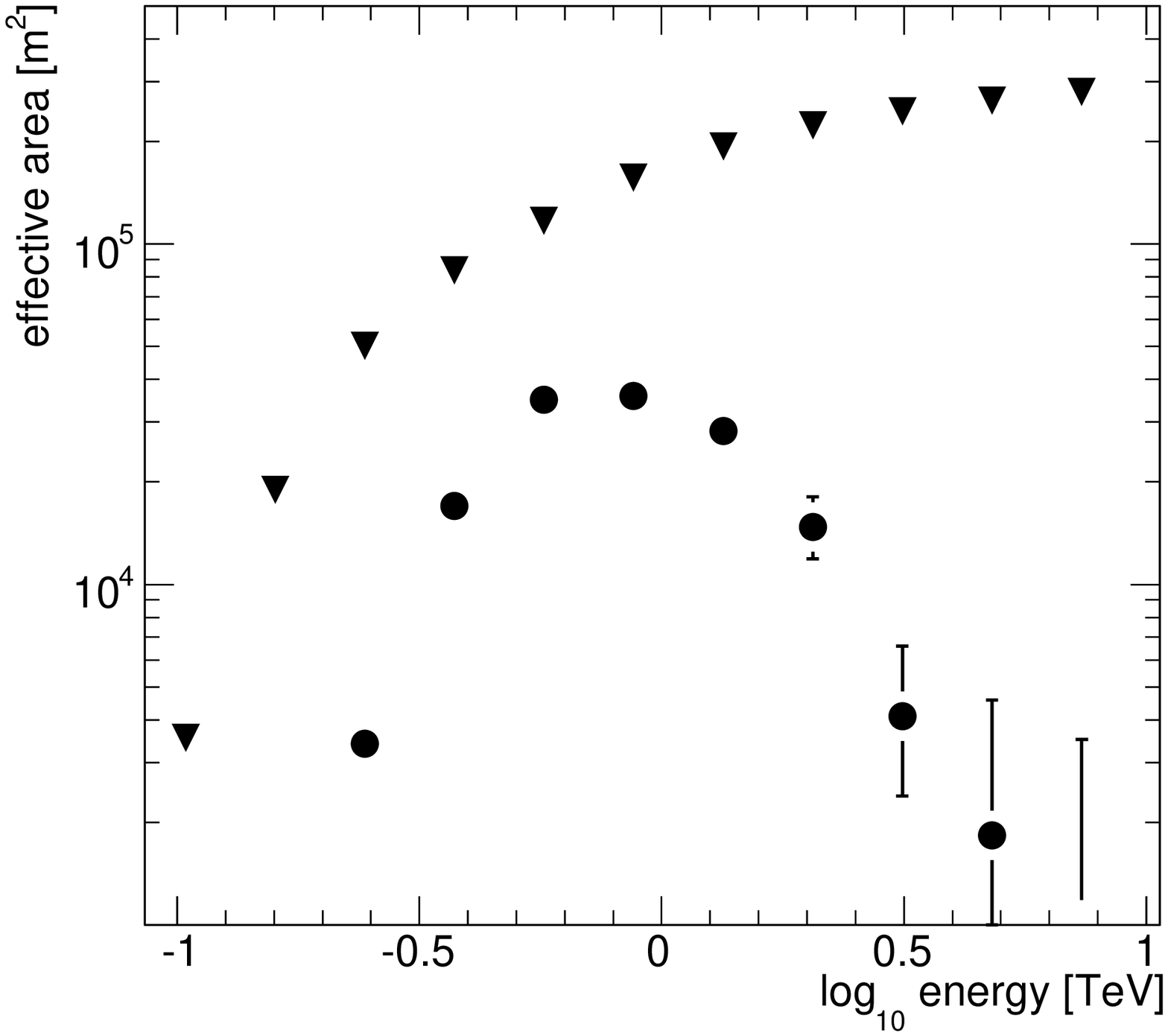}
\includegraphics*[width=0.41\textwidth]{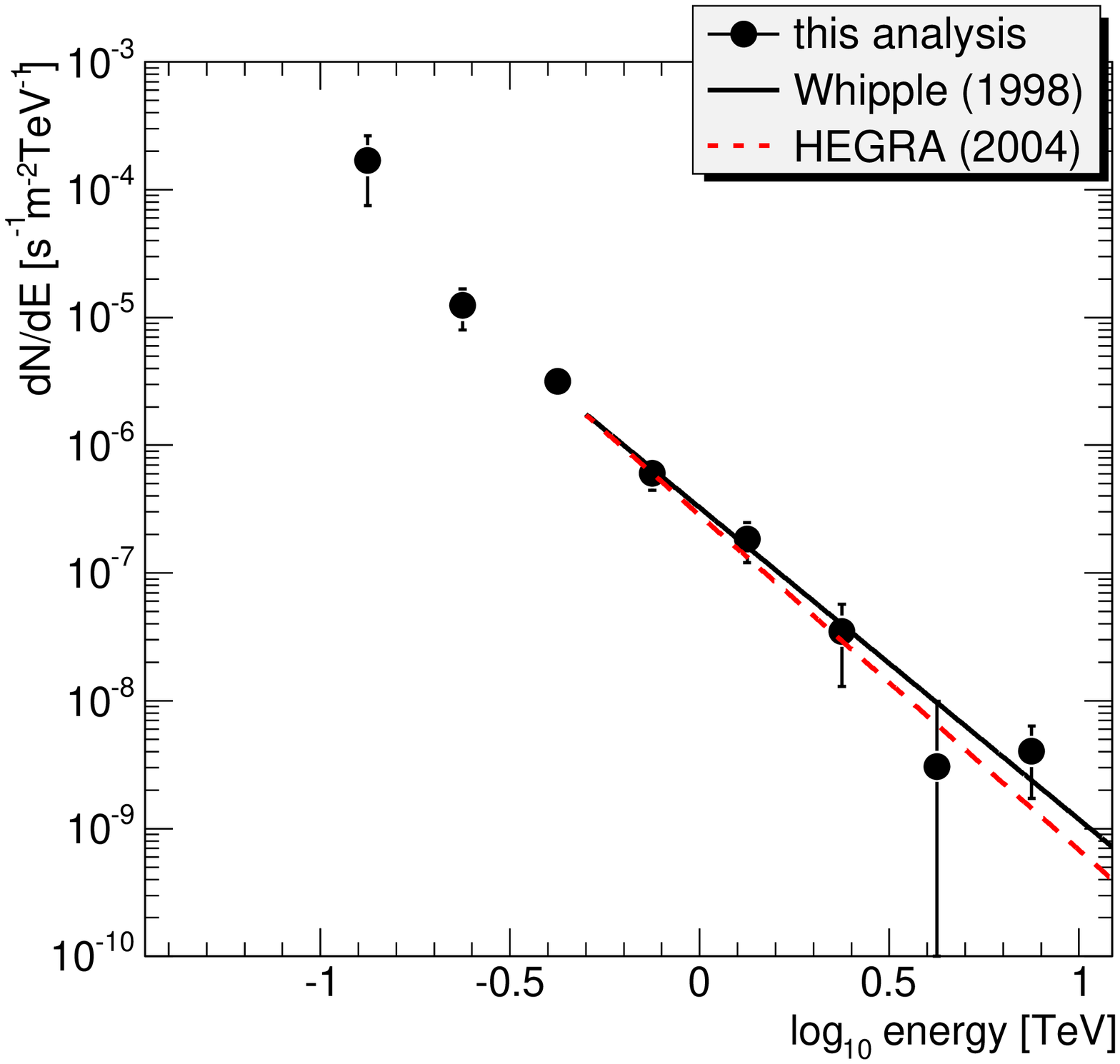}
\caption{ \label {fig2} Left: Effective area of Telescope 1 after $\gamma$-selection
cuts (filled circles) and extended supercuts (filled triangles).
Right: Energy spectrum of the Crab Nebula compared with earlier results from Whipple \cite{Hillas:1998}
and HEGRA \cite{Aharonian:2004}.}
\end{center}
\end{figure}

The effective area of Telescope 1 after cuts is shown in Fig.\ref{fig2} (left).
Most of the high-energy events fail the hard $\gamma$-selection cuts due to the cut on distance.
This is different for the application of extended supercuts \cite{Mohanty}, which do not result in a similar
good rejection of background events, but
take the dependence of the image parameters on image size into account.
The maximum effective area after extended supercuts is $\sim 2.8\cdot 10^5$ m$^{2}$.
The measured gamma ray rate from the observation of the Crab Nebula is $2.1\pm0.2$ $\gamma$'s/min.
Using the Crab spectra reported in \cite{Hillas:1998} results in a simulated gamma ray rate
from the Crab Nebula of 2.2 $\gamma$'s/min.
The energy threshold of a Cherenkov telescope is conventionally defined as the position of the
peak of the energy spectrum of the source convoluted with the effective area curve of the detector.
According to this definition, the threshold 
is 150 GeV at trigger level, 160 GeV after applying extended super cuts, and 370 GeV
after applying the hard cuts described earlier.
The right-hand side of Fig.\ref{fig2} shows the reconstructed energy spectrum of the Crab Nebula after
applying extended supercuts to the data set.
A power law fit to the data points results in a spectral index of $2.6\pm0.3$ and a flux constant of
$(3.26\pm0.9)\cdot10^{-7}$ m$^{-2}$s$^{-1}$TeV$^{-1}$ (statistical errors only).
This agrees well with earlier measurements by other telescopes.

The agreement of the Monte Carlo simulations with observational data shows that we have a good understanding
of the first VERITAS telescope and that the design performance is being met.
The final configuration of VERITAS with four telescopes will have a significant higher sensitivity and 
an energy threshold of 110 GeV \cite{Fegan:2003}.

\vspace{-0.35cm}
\subsection*{Acknowledgments}

This research is supported by grants from the U.S.~Department of Energy, the National Science Foundation,
the Smithsonian Institution, by NSERC in Canada, by Science Foundation Ireland and by PPARC in the UK.
Gernot Maier acknowledges the support as a Feodor Lynen Fellow of the Alexander von Humboldt foundation.


\vspace{-0.45cm}
\begin{thebibliography}{99}

\bibitem{JH05} J.~Holder et al., {\em Status and performance of the first VERITAS telescope}, 29th ICRC, Pune (2005)
\bibitem{Weekes02} T.~Weekes et al., Astroparticle Physics, 17, 221 (2002)
\bibitem{Hillas:1985} A.~M.~Hillas, Proc. 19th Int.~Cosmic Ray Conf., 3, 445 (1985) 
\bibitem{Heck} D.~Heck et al., Report FZKA 6019, Forschungszentrum Karls\-ruhe (1998)
\bibitem{LeBohec} C.Duke \& S.LeBohec, http://www.physics.utah.edu/gammaray/GrISU/
\bibitem{Hoerandel} J.~H{\"o}randel, Astroparticle Physics, 19, 193 (2003)
\bibitem{Modtran} Kneizys et al., {\em The Modtran 2/3 report and lowtran 7 model}, Technical Report,
Ontar Corporation (1996)
\bibitem{Mohanty} G.~Mohanty et al., Astroparticle Physics, 9, 15 (1998)
\bibitem{Hillas:1998} A.M.~Hillas et al.,The Astrophysical Journal, 503, 744 (1998)
\bibitem{Aharonian:2004} F.~Aharonian et al., The Astrophysical Journal, 614, 897 (2004)
\bibitem{Fegan:2003} S.~Fegan, Proc. 28th Int.~Cosmic Ray Conf., 2847 (2003)
\end{thebibliography}
\end{document}